\documentclass[preprint,showpacs,showkeys,preprintnumbers,amsmath,amssymb]{revtex4}
\usepackage{amssymb}
\usepackage{graphicx}
\usepackage{subfigure}
\usepackage{float}

\newcommand\abs[1]{\left\lvert #1 \right\rvert}

\begin{document}

\title{A nonlocal nonlinear Schr\"{o}dinger equation derived from a two-layer fluid model}

\author{ Xi-zhong Liu}
\affiliation{Institute of Nonlinear Science, Shaoxing University, Shaoxing 312000, China}
\begin{abstract}
By applying a simple symmetry reduction on a two-layer liquid model, a nonlocal counterpart of it is obtained. Then a general form of nonlocal nonlinear Schr\"{o}dinger (NNLS) equation with shifted parity, charge-conjugate and delayed time reversal is obtained by using multi-scale expansion method. Some kinds of elliptic periodic wave solutions of the NNLS equation are obtained by using function expansion method, which contain soliton solutions and kink solutions when the modulus taking as unity. Some representative figures of these solutions are given and analyzed in detail. In addition, by carrying out the classical symmetry method on the NNLS equation, not only the Lie symmetry group but also the related symmetry reduction solutions are given.
\end{abstract}

\pacs{02.30.Jr,\ 02.30.Ik,\ 05.45.Yv,\ 47.35.Fg}

\keywords{nonlocal nonlinear Schr\"{o}dinger equation, periodic waves, symmetry reduction solutions}

\maketitle

\section{Introduction}
In 2013, Ablowitz and Musslimani \cite{muss} introduced a PT symmetric nonlocal Schr\"{o}dinger (NNLS) equation
\begin{equation}\label{schr}
iq_{t}(x,t)=q_{xx}(x,t)\pm q(x,t)q^*(-x,t)q(x,t),
\end{equation}
with $\ast$ being complex conjugate and $q$ being a complex valued function of the real variables
$x$ and $t$. Eq. \eqref{schr} is an integrable infinite dimensional Hamiltonian equation, which can be solved by inverse scattering transform and possesses infinitely number of conservation laws. Soon thereafter interest in nonlocal nonlinear systems has grown significantly and many nonlocal systems have been constructed by making nonlocal symmetry reductions of some general AKNS scattering problems, such as the nonlocal derivative nonlinear Schr\"{o}dinger equation, nonlocal modified KdV equation, nonlocal sine-Gordon equation, nonlocal
Davey-Stewartson equation, nonlocal Ablowitz-Ladik equation, and so on \cite{markj,fokas,maly,yk}. At the same time many effective methods, such as Hirota's bilinear method, inverse scattering theory, Darboux transformations and so on,  are developed to find N-soliton solutions, conservation laws, rational solitons solutions, rogue waves, peakon solutions, etc., of nonlocal systems\cite{maly,yanzhenya,muss3,fokas,kh,louqiao}

Due to the abundant existence of nonlocal phenomenon in real nature \cite{mau,coc,per,con}, Lou investigated nonlocal systems from perspective of physics and introduced the concept of ``Alice-Bob" (AB) system to study two intrinsically correlated events $A$ and $B$, which linked each other by $B=\hat{f}A$ with $\hat{f}$ being a suitably chosen operator. By applying the ``$AB-BA$ equivalence principle" and ``$\hat{P}_s$-$\hat{T}_d$-$\hat{C}$ principle'', many AB-type nonlocal versions of celebrated physical nonlinear equations have been obtained and studied, such as KdV equation, mKdV equation, KP equation, Schr\"{o}dinger equation, etc. \cite{louab, jiaman,licongcong,louab2}. Moreover, various exact solutions of these nonlocal systems are investigated with different methods. For example, for the AB-type KdV equation in \cite{jiaman}, $P_sT_d$ invariant multiple soliton solutions are obtained from the known ones of the KdV equation by fixing some arbitrary parameters therein, while $P_sT_d$ symmetry breaking solutions are obtained from the solutions of a coupled KdV system. However, for some non-integrable nonlocal systems it is difficult to solve them exactly. In this case, multi-scale expansion method have proven to be a efficient tool to find approximate solutions of many important physical models, by reducing them into some familiar equations, often with good integrability \cite{tang,tangkdv,tang2}. It is well known that symmetry analysis plays an important role in simplifying and even completely solving complicated nonlinear problems \cite{olver}. Applying Lie's standard symmetry method on integrable nonlinear equations, not only symmetry group but also symmetry reduction solutions can be obtained in a systematic way \cite{boussnd,2burepj,cmabo}. So it is interesting to applying this method on nonlocal systems and probing the new features therein.

The paper is organized as follows. In Sect. \uppercase\expandafter{\romannumeral2}, at first, a nonlocal version of a two-layer liquid model is obtained by using ``$AB-BA$ equivalence principle", then a general nonlocal nonlinear Schr\"{o}dinger (NNLS) equation with shifted parity, charge conjugate and delayed time reversal is derived from it by using multi-scale method. In Sect. \uppercase\expandafter{\romannumeral3}, 6 types of elliptic wave solutions of the NNLS equation are obtained by using function expansion method, which reduce to  kink (anti-kink) or bright (dark) type soliton solutions when taking the modulus as unity. Meanwhile, some representative solutions of them are shown graphically and analyzed in detail.  In Sect. \uppercase\expandafter{\romannumeral4}, applying Lie's standard symmetry method on the NNLS equation, not only the symmetry group but also the symmetry reduction solutions are obtained. The last section devotes to a summary and discussion.

\section{derivation of a general nonlocal nonlinear Schr\"{o}dinger equation with shifted parity, charge conjugate and delayed time reversal }
In this section, we take a two-layer liquid model \cite{ped}
\begin{equation}\label{q1}
q_{1t}+J\{\psi_1,q_1\}+\beta\psi_{1x}=0,
\end{equation}
\begin{equation}\label{q2}
q_{2t}+J\{\psi_2,q_2\}+\beta\psi_{2x}=0,
\end{equation}
where
\begin{equation}\label{rq1}
q_1=\psi_{1xx}+\psi_{1yy}+F(\psi_2-\psi_1),
\end{equation}
\begin{equation}\label{rq2}
q_2=\psi_{2xx}+\psi_{2yy}+F(\psi_1-\psi_2),
\end{equation}
\begin{equation}\label{rJ}
J\{a,b\}=a_xb_y-b_xa_y,
\end{equation}
as a starting point to derive a general nonlocal nonlinear Schr\"{o}dinger equation. In Eqs. \eqref{q1}-\eqref{rq2}, $F$ is a small constant, indicating coupling strength between two layers of fluid; $\beta=\beta_0(L^2/U)$ with $\beta_0=(2\omega_0/a_0)\cos(\varphi_0)$, where $a_0,\,\omega_0,\,\varphi_0$ being the earth's radius, the angular frequency of the earth's rotation and the latitude, respectively, $U$ is the characteristic velocity scale and $L$ is characteristic horizontal length scale. In derivation of Eqs. \eqref{q1}-\eqref{q2} in Ref. \cite{ped}, the constants are fixed as $L=10^6 m$ and $U=10^{-1} m s^{-1}$.

The nonlocal version of the two-layer liquid model \eqref{q1}-\eqref{q2} can be easily obtained by applying the so called``AB-BA equivalence principle" and ``$\hat{P}_s$-$\hat{T}_d$-$\hat{C}$ principle", i.e. by making the following symmetry constraint
\begin{equation}\label{pt1}
\psi_2=\hat{f}\psi_{1}\equiv\hat{P}_s^x\hat{T}_dC\psi_{1}=\psi_1^*(-x+x_0,y,-t+t_0).
\end{equation}

In order to derive a nonlocal nonlinear Schr\"{o}dinger (NNLS) equation, under the long wave approximation assumption in $x$-direction, the stream function is assumed as
\begin{equation}\label{psi1e}
\psi_1=\psi_1(x,y,t,\xi,\tau)=\psi_1(x,y,t,\epsilon(x-c_0t),\epsilon^2t),
\end{equation}
where $\epsilon$ is a small parameter and $c_0$ is an arbitrary constant. Here, we change the form of $\psi_1$ as
\begin{equation}\label{epsi1}
\psi_1=c_1+m_0y+\psi_{11}(x,y,t,\xi,\tau),
\end{equation}
with $c_1$ and $m_0$ being arbitrary constants and the last term can be expanded as
\begin{equation}\label{epsi11}
\psi_{11}(x,y,t,\xi,\tau)=\epsilon\phi_{11}+\epsilon^2\phi_{12}+\epsilon^3\phi_{13}+O(\epsilon^4),
\end{equation}
with $\phi_{1i}\equiv\phi_{1,i}(x,y,t,\xi,\tau),\,(i=1,2,3)$ being functions of indicated variables, while $F$ in the two-layer model can be taken as
\begin{equation}\label{ec}
F=F_0\epsilon^2,
\end{equation}
which is in consideration of the far weak coupling between two layers of the fluid.

Substituting Eq. \eqref{psi1e} with Eqs. \eqref{epsi1}, \eqref{epsi11} and \eqref{ec} into Eqs. \eqref{q1} and \eqref{q2} with \eqref{rq1}, \eqref{rq2}, \eqref{rJ} and vanishing coefficients of $O(\epsilon)$, we obtain
\begin{equation}\label{ep11}
m_0(\phi_{11yyx}+\phi_{11xxx})-\beta\phi_{11x}-(\phi_{11yy}+\phi_{11xx})_t=0,
\end{equation}
and
\begin{equation}\label{ep12}
m_0(\phi_{21yyx}+\phi_{21xxx})-\beta\phi_{21x}-(\phi_{21yy}+\phi_{21xx})_t=0,
\end{equation}
where $\phi_{21}=\hat{P}_s^{x}\hat{P}_s^{\xi}\hat{T}_d^{t}\hat{T}_d^{\tau}\hat{C}\phi_{11}=\phi_{11}^*(-x+x_0,y,-t+t_0,-\xi+\xi_0,-\tau+\tau_0)$.

The solutions of Eqs. \eqref{ep11} and \eqref{ep12} can be supposed having the form
\begin{equation}\label{sophi11}
 \phi_{11} = G_{0}(y)A(\xi, \tau)e^{i[k(x-\frac{x_0}{2})-\omega(t-\frac{t_0}{2})]}+c.c.\equiv G_{0}Ae^{i[k(x-\frac{x_0}{2})-\omega(t-\frac{t_0}{2})]}+c.c.,
\end{equation}
and
\begin{equation}\label{sophi21}
 \phi_{21} = G_{0}(y)B(\xi, \tau)e^{i[k(x-\frac{x_0}{2})-\omega(t-\frac{t_0}{2})]}+c.c.\equiv G_{0}Be^{i[k(x-\frac{x_0}{2})-\omega(t-\frac{t_0}{2})]}+c.c.,
\end{equation}
with $B(\xi,\tau)=\hat{P}_s^{\xi}\hat{T}_d^{\tau}\hat{C}A=A^*(-\xi+\xi0,-\tau+\tau_0)$ , where ``$c.c.$ " means complex conjugation of the previous terms. Now, by substituting Eqs. \eqref{sophi11} and \eqref{sophi21} into Eqs. \eqref{ep11} and \eqref{ep12}, we have
\begin{equation}\label{g0yy}
G_{0yy}= \frac{G_0k(m_0k^2+k\omega+\beta)}{m_0 k+\omega},
\end{equation}
which has a general solution
\begin{equation}
G_0=m_2\sin\big[y\sqrt{\frac{-k\beta}{km_0+\omega}-k^2}+m_1\big],
\end{equation}
with arbitrary constants $m_1$ and $m_2$.

Next, vanishing the coefficients of $O(\epsilon^2)$ and $O(\epsilon^3)$, respectively, lead to
\begin{multline}\label{eps21}
(\phi_{12x}+\phi_{11\xi})\beta-c_0(\phi_{11yy\xi}+\phi_{11 xx\xi})-m_0(3\phi_{11 xx\xi}+\phi_{11yy\xi}+\phi_{12yyx}+\phi_{12xxx})+\phi_{11x}\phi_{11yxx}\\+\phi_{11x}\phi_{11yyy}-\phi_{11yyx}\phi_{11y}-\phi_{11xxx}
\phi_{11y}+\phi_{12yyt}+2\phi_{11\xi xt}+\phi_{12xxt}=0,
\end{multline}

\begin{multline}\label{eps22}
(\phi_{22x}+\phi_{21\xi})\beta-c_0(\phi_{21yy\xi}+\phi_{21 xx\xi})-m_0(3\phi_{21 xx\xi}+\phi_{21yy\xi}+\phi_{22yyx}+\phi_{22xxx})+\phi_{21x}\phi_{21yxx}\\+\phi_{21x}\phi_{21yyy}
-\phi_{21yyx}\phi_{21y}-\phi_{21xxx}
\phi_{21y}+\phi_{22yyt}+2\phi_{21\xi xt}+\phi_{22xxt}=0,
\end{multline}
and
\begin{multline}\label{eps31}
(\phi_{12\xi}+\phi_{13x})\beta-(\phi_{12yy\xi}+2\phi_{11\xi\xi x}+\phi_{12\xi xx})c_0+(F_0\phi_{21x}+F_0\phi_{11x}-\phi_{13yyx}-\phi_{12yy\xi}-\phi_{13xxx}\\-3\phi_{11\xi\xi x}-3\phi_{12\xi xx})m_0-F_0\phi_{11t}-F_0\phi_{21t}+\phi_{12x}\phi_{11yyy}+\phi_{12x}\phi_{11yxx}+\phi_{11\xi}\phi_{
11yyy}+\phi_{11\xi}\phi_{11yxx}\\+\phi_{11x}\phi_{12yxx}-\phi_{11y}\phi_{12yyx}-\phi_{11y}\phi_{11yy\xi}
-\phi_{11y}\phi_{12xxx}-3\phi_{11y}\phi_{11\xi xx}-\phi_{12y}\phi_{11yyx}-\phi_{12y}\phi_{11xxx}\\+\phi_{11x}\phi_{12yyy}+2\phi_{11x}\phi_{11y\xi x}+\phi_{11xx\tau}+2\phi_{12\xi xt}+\phi_{13yyt}+\phi_{11yy\tau}+\phi_{11\xi\xi t}+\phi_{13xxt}=0,
\end{multline}
\begin{multline}
(\phi_{22\xi}+\phi_{23x})\beta-(\phi_{22yy\xi}+2\phi_{21\xi\xi x}+\phi_{22\xi xx})c_0+(F_0\phi_{11x}+F_0\phi_{21x}-\phi_{23yyx}-\phi_{22yy\xi}-\phi_{23xxx}\\-3\phi_{21\xi\xi x}-3\phi_{22\xi xx})m_0-F_0\phi_{21t}-F_0\phi_{21t}+\phi_{22x}\phi_{21yyy}+\phi_{22x}\phi_{21yxx}+\phi_{21\xi}\phi_{
21yyy}+\phi_{21\xi}\phi_{21yxx}\\+\phi_{21x}\phi_{22yxx}-\phi_{21y}\phi_{22yyx}-\phi_{21y}\phi_{21yy\xi}
-\phi_{21y}\phi_{22xxx}-3\phi_{21y}\phi_{21\xi xx}-\phi_{22y}\phi_{21yyx}-\phi_{22y}\phi_{21xxx}\\+\phi_{21x}\phi_{22yyy}+2\phi_{21x}\phi_{21y\xi x}+\phi_{21xx\tau}+2\phi_{22\xi xt}+\phi_{23yyt}+\phi_{21yy\tau}+\phi_{21\xi\xi t}+\phi_{23xxt}=0,
\end{multline}

where $\phi_{2i}=\hat{P}_s^{x}\hat{P}_s^{\xi}\hat{T}_d^{t}\hat{T}_d^{\tau}\hat{C}\phi_{1i}
=\phi_{1i}^*(-x+x_0,y,-t+t_0,-\xi+\xi_0,-\tau+\tau_0),\,(i=2,\,3)$.

It can be verified that $\phi_{12}$ and $\phi_{22}$ in Eqs. \eqref{eps21} and \eqref{eps22} have the following form
\begin{equation}\label{exphi12}
\phi_{12} =(G_1B+G_2A+iG_3A_{\xi})e^{i[k(x-\frac{x_0}{2})-\omega(t-\frac{t_0}{2})]}+c.c.+G_4\left|A\right|^2
+G_5\left|B\right|^2+G_6(AB^*+A^*B),
\end{equation}
and
\begin{equation}\label{exphi22}
\phi_{22}=(G_1A+G_2B-iG_3B_{\xi})
e^{i[k(x-\frac{x_0}{2})-\omega(t-\frac{t_0}{2})]}+c.c.+G_4\left|B\right|^2
+G_5\left|A\right|^2+G_6(AB^*+A^*B),
\end{equation}
where $G_i\, (i=1,\,2,\, 3)$ are determined by
\begin{subequations}\label{G123}
\begin{eqnarray}
G_{1yy}&=& \frac{G_1k(m_0k^2+k\omega+\beta)}{m_0 k+\omega},\\
G_{2yy}&=& \frac{G_2k(m_0k^2+k\omega+\beta)}{m_0 k+\omega},\\
G_{3yy}&=& \frac{[(c_0k-\omega)\beta-2k(m_0k+\omega)^2]G_0+k(m_0k+\omega)(m_0k^2+k\omega+\beta)G_3}{(m_0 k+\omega)^2}.
\end{eqnarray}
\end{subequations}
Substituting Eqs. \eqref{exphi12}, \eqref{exphi22} with Eq. \eqref{G123} into Eq. \eqref{eps31} and setting $\phi_{13}=\phi_{23}=0$, we get
\begin{equation}\label{gama10}
\gamma_1e^{i[k(x-\frac{x_0}{2})-\omega(t-\frac{t_0}{2})]}+\gamma_1^*e^{-i[k(x-\frac{x_0}{2})
-\omega(t-\frac{t_0}{2})]}+\gamma_0=0,
\end{equation}
where
\begin{multline}\label{gama1}
\gamma_1=G_0\beta k(m_0k+\omega)A_{\tau}+(m_0k+\omega)(2m_0^2k^3+4m_0k^2\omega-\beta c_0k+2k\omega^2+\beta\omega)(
A_{\xi}G_2+B_{\xi}G_1)\\-i\big[G_{4y}\left|A\right|^2+AB^*G_{6y}+A^*B G_{6y}+G_{5y}\left|B\right|^2\big]AG_0\beta(m_0k+\omega)k^2
+\big\{2iG_3m_0^3k^4\\+im_0^2(6G_3\omega-G_0m_0)k^3
-im_0(G_3\beta c_0-6G_3\omega^2+3G_0m_0\omega)k^2+i[2G_3\omega^3
-G_3\beta(c_0-m_0)\omega\\-G_0\beta c_0(c_0+m_0)-3G_0m_0\omega^2]k+i\omega(G_3\beta\omega+2G_0\beta c_0-G_0\omega^2)\big\}A_{\xi\xi}+i\big[G_{4yyy}\left|A\right|^2+AB^*G_{6yyy}\\+A^*B G_{6yyy}+G_{5yyy}\left|B\right|^2\big]
AG_0k(m_0k+\omega)^2+i F_0(A+B)G_0(m_0k+\omega)^3,
\end{multline}
and
\begin{multline}
\gamma_0=\big\{[2G_0\beta k(c_0k-\omega)G_{0y}-(km_0+\omega)^2(c_0+m_0)G_{4yy}+G_4\beta(km_0+\omega)^2]A^*\\-(km_0+\omega)^2[
(c_0+m_0)G_{6yy}-G_6\beta]B^*\big\}A^*_{\xi}+\big\{[2G_0\beta k(c_0k-\omega)G_{0y}-(km_0+\omega)^2(c_0+m_0)G_{4yy}\\+G_4\beta(km_0+\omega)^2]A-(km_0+\omega)^2
(G_{6yy}c_0+G_{6yy}m_0-G_6\beta)B\big\}A^*_{\xi}-(km_0+\omega)^2\big\{(G_{6yy}c_0+G_{6yy}m_0\\-G_6\beta)A^*
+(G_{5yy}c_0+G_{5yy}m_0-G_5\beta) B^*\big\}B_{\xi}-(km_0+\omega)^2(G_{6yy}c_0+G_{6yy}m_0-G_6\beta) B^*_{\xi}A\\-(km_0+\omega)^2(G_{5yy}c_0+G_{5yy}m_0-G_5\beta) B^*_{\xi}B.
\end{multline}
It is obvious that the coefficients of different powers of exponentials in Eq. \eqref{gama10} must to be zero. As for $\gamma_0=0$, it leads to
\begin{subequations}\label{G456}
\begin{eqnarray}
G_{4yy}&=& \frac{\beta[2kG_{0y}(c_0k-\omega)G_0+(m_0k+\omega)^2G_4]}{(c_0+m_0)(m_0 k+\omega)^2},\\
G_{5yy}&=& \frac{\beta G_5}{m_0+c_0},\\
G_{6yy}&=& \frac{\beta G_6}{m_0+c_0}.
\end{eqnarray}
\end{subequations}
Substituting Eq. \eqref{G456} into Eq. \eqref{gama1} and setting it to zero, we finally get the desired NNLS equation
\begin{equation}\label{nonnls}
iA_{\tau}+e_1A_{\xi\xi}+ie_2A_{\xi}+ie_3B_{\xi}+e_4\left|A\right|^2A+e_5B^*A^2+e_6(\left|A\right|^2B
+\left|B\right|^2A)+e_7A+e_7B=0,
\end{equation}
with
\begin{equation}\label{nonnlsB}
B(\xi,\tau)=\hat{P}_s^{\xi}\hat{T}_d^{\tau}\hat{C}A,
\end{equation}
where the constants $e_i\,(i=1\cdots7)$ are determined by
\begin{multline}
e_1=-\frac{1}{y_0(m_0k+\omega)\beta k}\int_0^{y_0}\frac{ 1}{G_0}[(m_0k+\omega)^3(2G_3k-G_0)-(c_0k-\omega)
(G_3m_0k+G_3\omega\\+G_0c_0+G_0m_0)\beta]dy,
\end{multline}
\begin{equation}
e_2=\frac{1}{y_0\beta k}\int_0^{y_0}\frac{1}{G_0}[2k(km_0+\omega)^2-(c_0k-\omega)\beta]G_2dy,
\end{equation}
\begin{equation}
e_3=\frac{1}{y_0\beta k}\int_0^{y_0}\frac{1}{G_0}[2k(km_0+\omega)^2-(c_0k-\omega)\beta]G_1dy,
\end{equation}
\begin{multline}
e_4=\frac{(c_0k-\omega)}{(c_0+m_0)y_0(m_0 k+\omega)^2}\int_0^{y_0}\big[-2k(km_0+\omega)G_{0y}^2+(km_0+\omega)^2G_{4y}\\-2k^2(k^2m_0+k\omega
+\beta)G_0^2\big]dy,
\end{multline}
\begin{equation}
e_5=\frac{1}{(c_0+m_0)y_0}\int_0^{y_0}(c_0k-\omega)G_{6y}dy,
\end{equation}
\begin{equation}
e_6=\frac{1}{(c_0+m_0)y_0}\int_0^{y_0}(c_0k-\omega)G_{5y}dy,
\end{equation}
\begin{equation}
e_7=-\frac{F_0(m_0k+\omega)^2}{\beta k}.
\end{equation}
\section{periodic wave solutions of the nonlocal nonlinear Schr\"{o}dinger equation}
As well known, function expansion method is efficient in obtaining exact solutions for nonlinear equations. As for the NNLS equation \eqref{nonnls}, we use elliptic function expansion method to get the following 6 types of periodic wave solutions.

\textbf{Solution 1} The first periodic wave solution is
\begin{equation}\label{snsol}
A=\pm mk\sqrt{\frac{2 e_1}{e_5-e_4}} {\rm SN}e^{i[k_1(\xi-\frac{\xi_0}{2})+\omega_1(\tau-\frac{\tau_0}{2})]},
\end{equation}
with ${\rm SN}\equiv {\rm sn}[k(\xi-\frac{\xi_0}{2})+\omega(\tau-\frac{\tau_0}{2}),m]$, $m$ being modulus of the Jacobi elliptic ${\rm sn}$ function, and
\begin{equation}\label{rb2}
k_1=-\frac{e_2k-e_3k+\omega}{2e_1k},\,\omega_1=-\frac{k^2[4e_1^2k^2(m^2+1)-(e_2-e_3)^2]
+\omega^2}{4e_1k^2}.
\end{equation}

\textbf{Solution 2} The second periodic wave solution is
\begin{equation}\label{cnsol}
A = \pm mk\sqrt{\frac{2e_1}{e_4+e_5+2e_6}}{\rm CN}e^{i[k_1(\xi-\frac{\xi_0}{2})+\omega_1(\tau-\frac{\tau_0}{2})]},
\end{equation}
with ${\rm CN}\equiv {\rm cn}[k(\xi-\frac{\xi_0}{2})+\omega(\tau-\frac{\tau_0}{2}),m]$, $m$ being modulus of the Jacobi elliptic ${\rm cn}$ function, and
\begin{equation}\label{rb2}
k_1=-\frac{e_2k+e_3k+\omega}{2e_1k},\,\omega_1=\frac{k^2[4e_1^2k^2(2m^2-1)+8e_1e_7
+(e_2+e_3)^2]-\omega^2}{4e_1k^2}.
\end{equation}

\textbf{Solution 3} The third periodic wave solution is
\begin{equation}\label{cnsol}
A = \pm k\sqrt{\frac{2e_1}{e_4+e_5+2e_6}}{\rm DN}e^{i[k_1(\xi-\frac{\xi_0}{2})+\omega_1(\tau-\frac{\tau_0}{2})]},
\end{equation}
with ${\rm DN}\equiv {\rm dn}[k(\xi-\frac{\xi_0}{2})+\omega(\tau-\frac{\tau_0}{2}),m]$, $m$ being modulus of the Jacobi elliptic ${\rm dn}$ function, and
\begin{equation}\label{rb2}
k_1=-\frac{e_2k+e_3k+\omega}{2e_1k},\,\omega_1=-\frac{k^2[4e_1^2k^2(m^2-2)-8e_1e_7
-(e_2+e_3)^2]+\omega^2}{4e_1k^2}.
\end{equation}

\textbf{Solution 4} The fourth periodic wave solution is
\begin{equation}\label{sol4}
A = \pm k\sqrt{\frac{e_1}{2(e_4+e_5+2e_6)}}({\rm DN}\pm m{\rm CN})e^{i[k_1(\xi-\frac{\xi_0}{2})+\omega_1(\tau-\frac{\tau_0}{2})]},
\end{equation}
with
\begin{equation}\label{rb2}
k_1=-\frac{e_2k+e_3k+\omega}{2e_1k},\,\omega_1=\frac{k^2[2e_1^2k^2(m^2+1)+8e_1e_7+(e_2+e_3)^2]
-\omega^2}{4e_1k^2}.
\end{equation}

\textbf{Solution 5} Under the condition $e_3=0$ and $e_4=e_6$, the fifth periodic wave solution is
\begin{equation}\label{cnsol}
A = \frac{2m\sqrt{-(e_5m^2+3e_6m^2+e_5-e_6)e_7}}{e_5m^2+3e_6m^2+e_5-e_6}({\rm DN}+ {\rm SN})e^{i[k_1(\xi-\frac{\xi_0}{2})+\omega_1(\tau-\frac{\tau_0}{2})]},
\end{equation}
with
\begin{equation}\label{rb2}
k_1=-\frac{e_1\omega+e_2\sqrt{-2e_1e_7}}{2e_1\sqrt{-2e_1e_7}},\,
k=\frac{\sqrt{-2e_1e_7}}{e_1},
\end{equation}
and
\begin{equation}
\omega_1=
\frac{2e_7(e_5m^4+3e_6m^4-4e_6m^2+e_5-e_6)}{e_5m^2+3e_6m^2+e_5-e_6}+\frac{\omega^2}{8e_7}
+\frac{e_2^2}{4e_1}.
\end{equation}
\textbf{Solution 6} Under the condition $e_3=0$ and $e_4=e_6$, the sixth periodic wave solution reads
\begin{equation}\label{sncnsol}
A =\sqrt{\frac{-2e_7}{e_5+e_6}}({\rm SN}+ {\rm CN})e^{i[k_1(\xi-\frac{\xi_0}{2})+\omega_1(\tau-\frac{\tau_0}{2})]},
\end{equation}
with
\begin{equation}\label{rb2}
k_1=-\frac{e_1m\omega+e_2\sqrt{-2e_1e_7}}{2e_1\sqrt{-2e_1e_7}},\,
k=\frac{\sqrt{-2e_1e_7}}{e_1m},
\end{equation}
and
\begin{equation}
\omega_1=\frac{(e_5+e_6)e_1(m^4\omega^2+16e_7^2)-2e_7(16e_1e_6e_7-e_2^2e_5-e_2^2e_6)m^2}
{8e_1e_7m^2(e_5+e_6)}.
\end{equation}

Here are some remarks on the above solutions. First, it is remarked that unlike the solutions of 5 and 6, the solutions of 1-4 are obtained without any constraints on the coefficients of the NNLS equation, which can have more potential applications in real physics. Second, from the odd or even property of Jacobi elliptic functions, the solutions of 1, 5, 6 are $\hat{P}_s^{\xi}\hat{T}_d^{\tau}\hat{C}$ symmetry breaking, while the other ones, i.e. solutions of 2, 3, 4, are $\hat{P}_s^{\xi}\hat{T}_d^{\tau}\hat{C}$ symmetry invariants. Third, when the modulus $m$ taking as unity, the solutions of 1 becomes kink or anti-kink soliton, the solutions of 2, 3, 4 become bright or dark solitons, while solutions of 5 and 6 are interaction solutions between these two kinds of solitons.

\begin{figure}
\centering
\subfigure[]{
\label{3dsol4a} %% label for first subfigure
\includegraphics[width=0.4\textwidth]{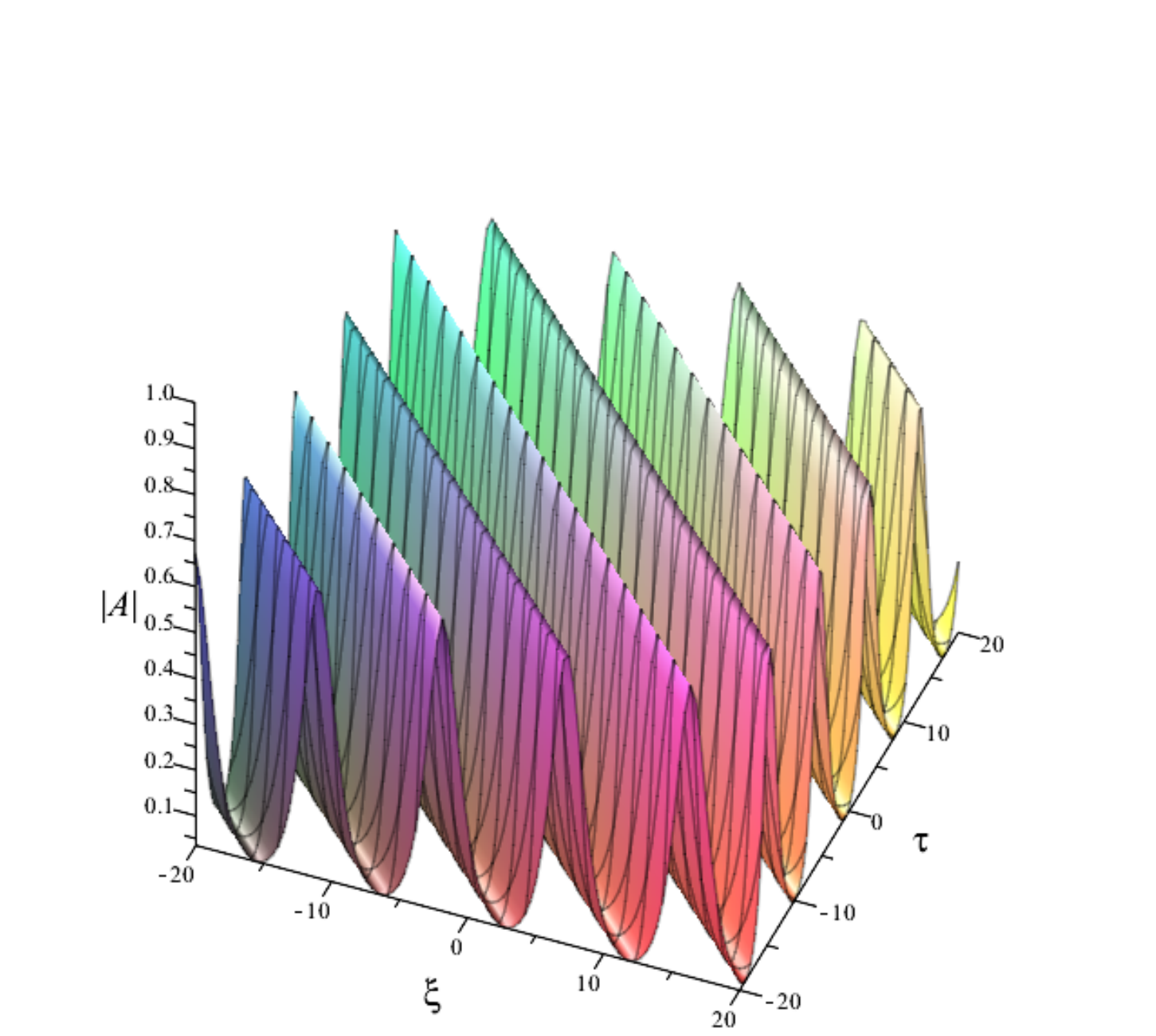}}
\subfigure[]{
\label{3dsol4b} %% label for first subfigure
\includegraphics[width=0.4\textwidth]{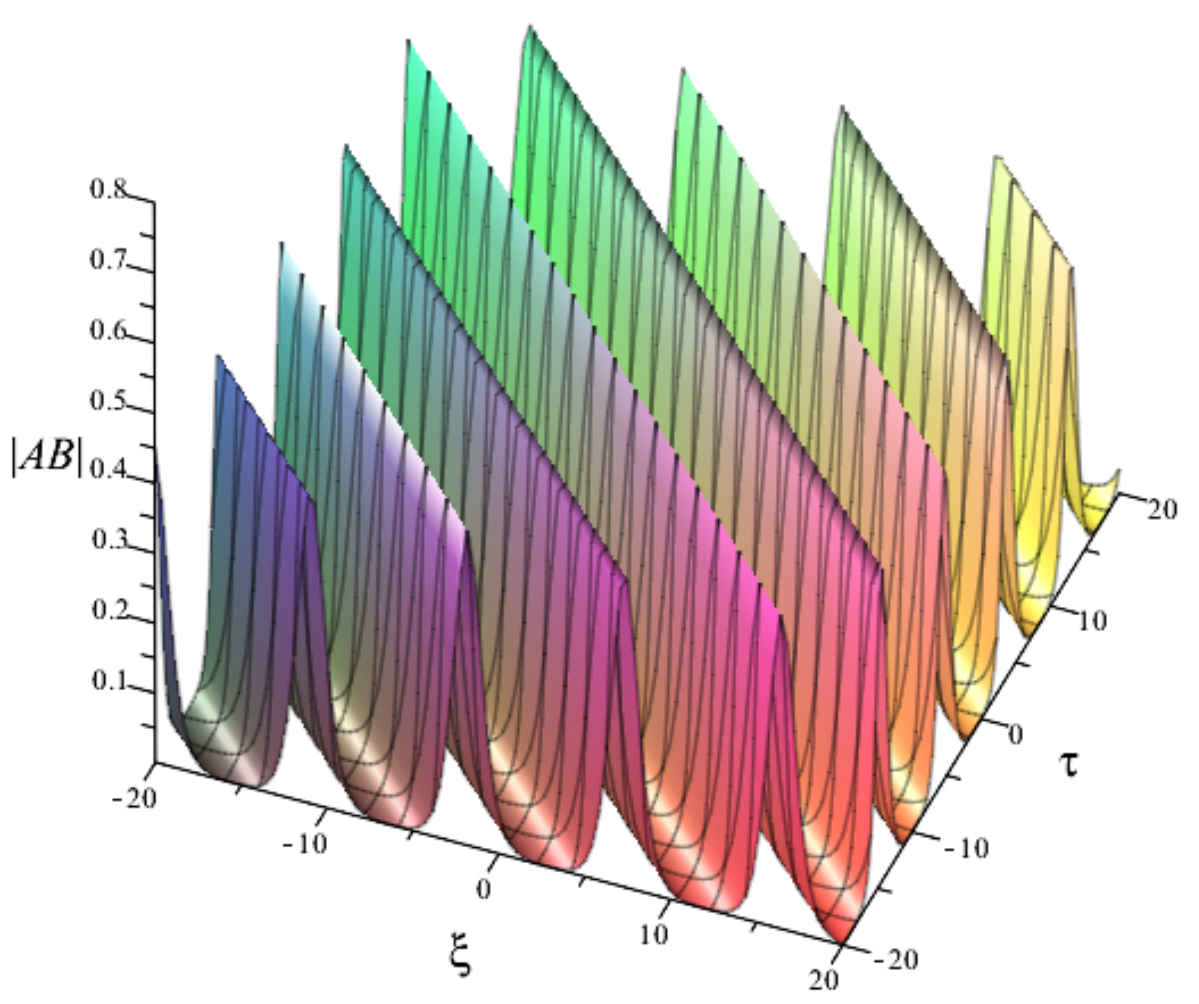}}
\caption{The three dimensional structure of the solution 4 with the parameters in Eq. \eqref{parasol4} for the quantity $\abs{A}$ (left) and $\abs{AB}$ (right).}
\label{3dsol4} %% label for entire figure
\end{figure}

\begin{figure}
\centering
\subfigure[]{
\label{1dsol4a} %% label for first subfigure
\includegraphics[width=0.4\textwidth]{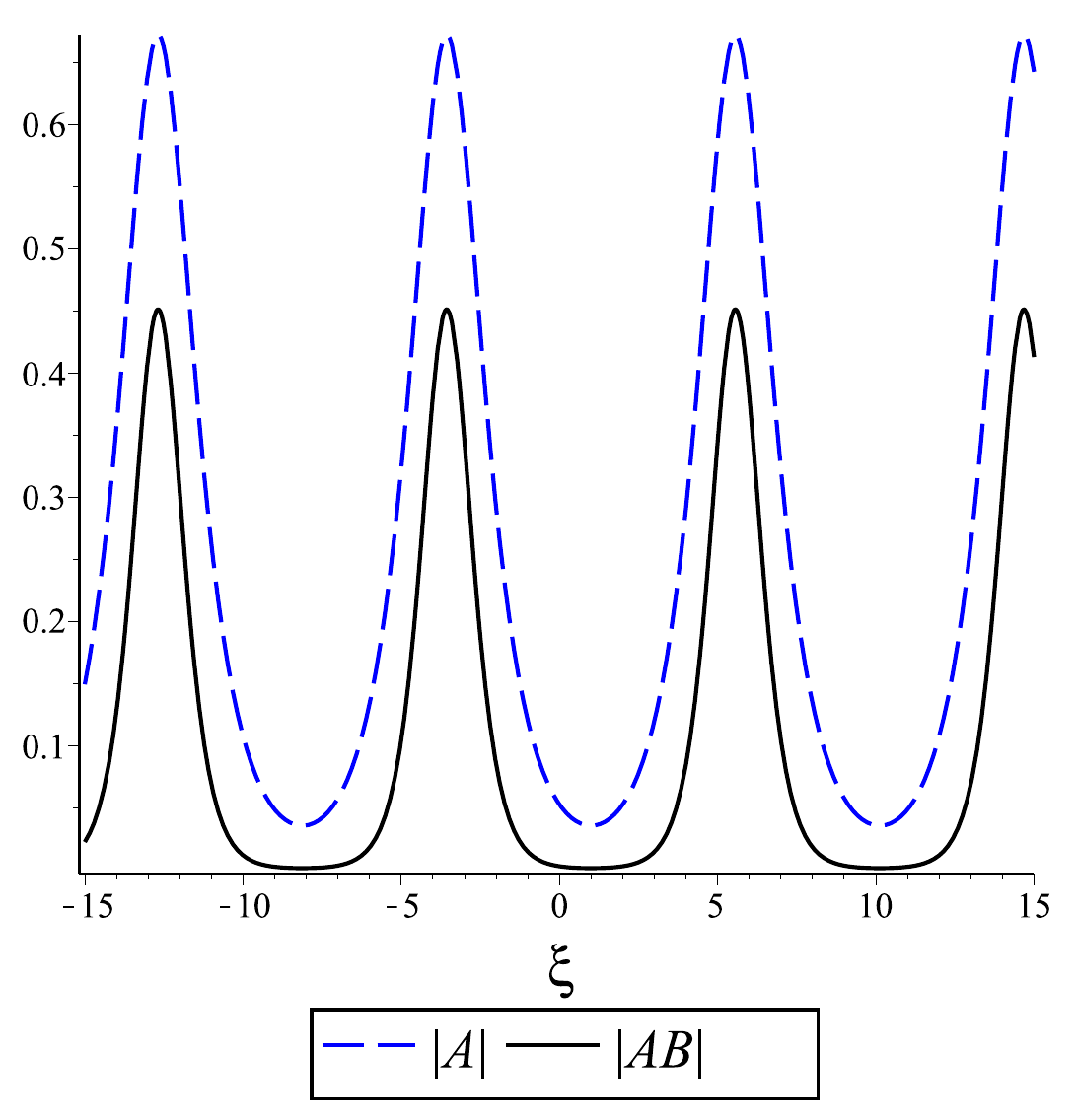}}
\subfigure[]{
\label{1dsol4b} %% label for first subfigure
\includegraphics[width=0.4\textwidth]{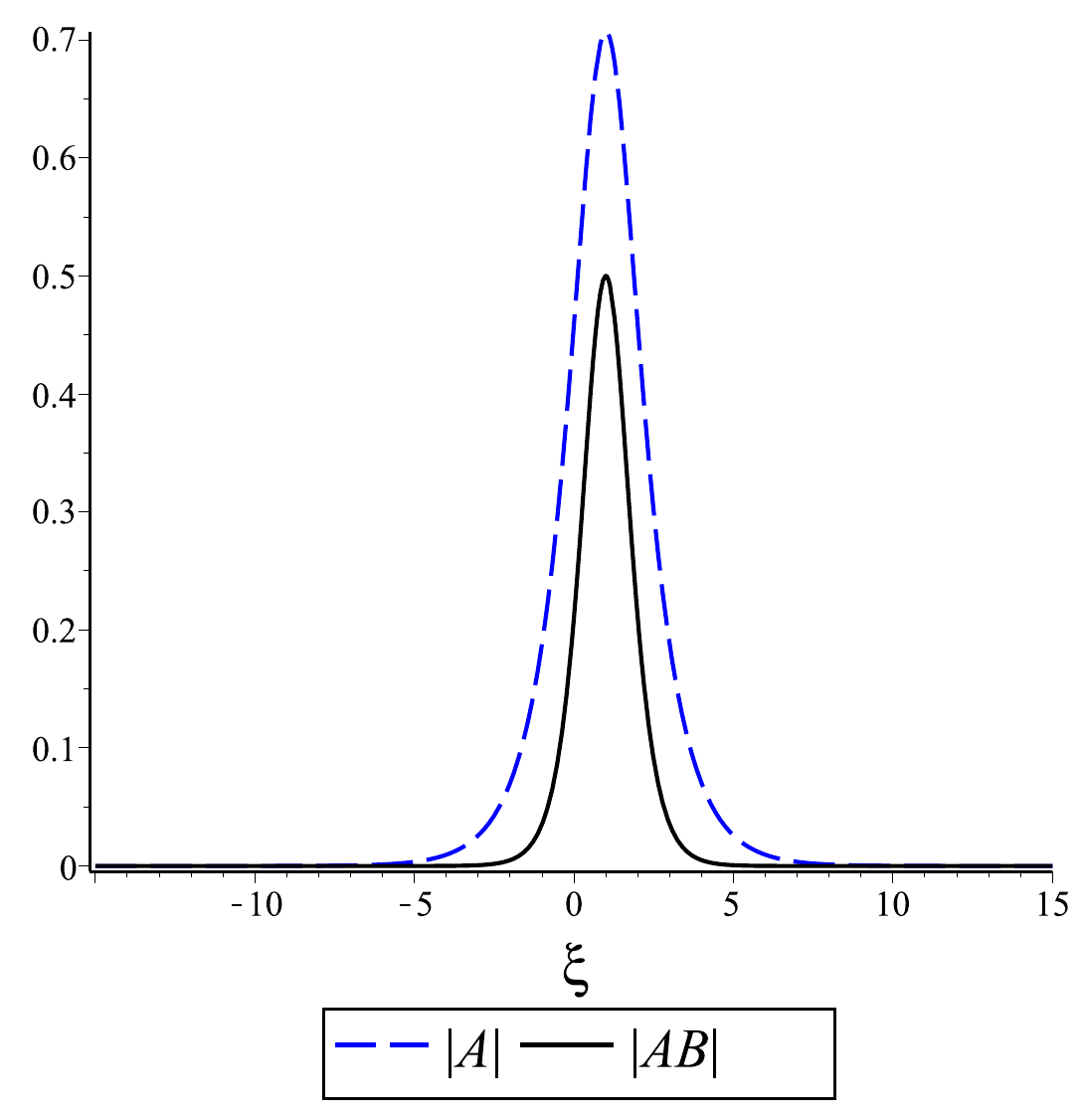}}
\caption{The one dimensional plot of the solution 4: (a) the wave patterns of $\abs{A}$ (blue-dash line) and $\abs{AB}$ (black-solid line) with the parameters in Eq. \eqref{parasol4} and $\tau=0$; (b) when the parameters being fixed as $e_1=e_2=e_4=e_5=e_6=k=\omega=\xi_0=\tau_0=1,\,e_3 = 0,\,e_7 = -1, m = 1$ the structure of $\abs{A}$ (blue-dash line) and $\abs{AB}$ (black-solid line).}
\label{inden} %% label for entire figure
\end{figure}

To show some special features of these solutions, we show some representative ones graphically.  Fig. \ref{3dsol4a}, \ref{3dsol4b} show three dimensional structure of $\abs{A}$ and $\abs{AB}$ of the solution 4, respectively, where the parameters are fixed as
\begin{equation}\label{parasol4}
e_1=e_2=e_4=e_5=e_6=k=\omega=\xi_0=\tau_0=1,\,e_3 = 0,\,e_7 = -1, m = 0.9,
\end{equation}
while Fig. \ref{1dsol4a} shows them both one dimensionally at $\tau=0$. When taking $m=1$ and other parameters remain unchanged as in Eq. \eqref{parasol4}, the quantities $\abs{A}$ and $\abs{AB}$ of the solution 4 become bright solitons, both of which are depicted one dimensionally in Fig. \ref{1dsol4b}. It can be seen from Fig. \eqref{3dsol4} that $\abs{A}$ and $\abs{AB}$ are all periodic waves, but Fig. \ref{1dsol4a} shows that the latter one has  a longer period but smaller amplitude than the former one. As for the bright solitons, Fig. \ref{1dsol4b} shows that the amplitude of soliton $\abs{A}$ is bigger than the one of soliton $\abs{AB}$.

\begin{figure}
\centering
\subfigure[]{
\label{3dsol5a} %% label for first subfigure
\includegraphics[width=0.4\textwidth]{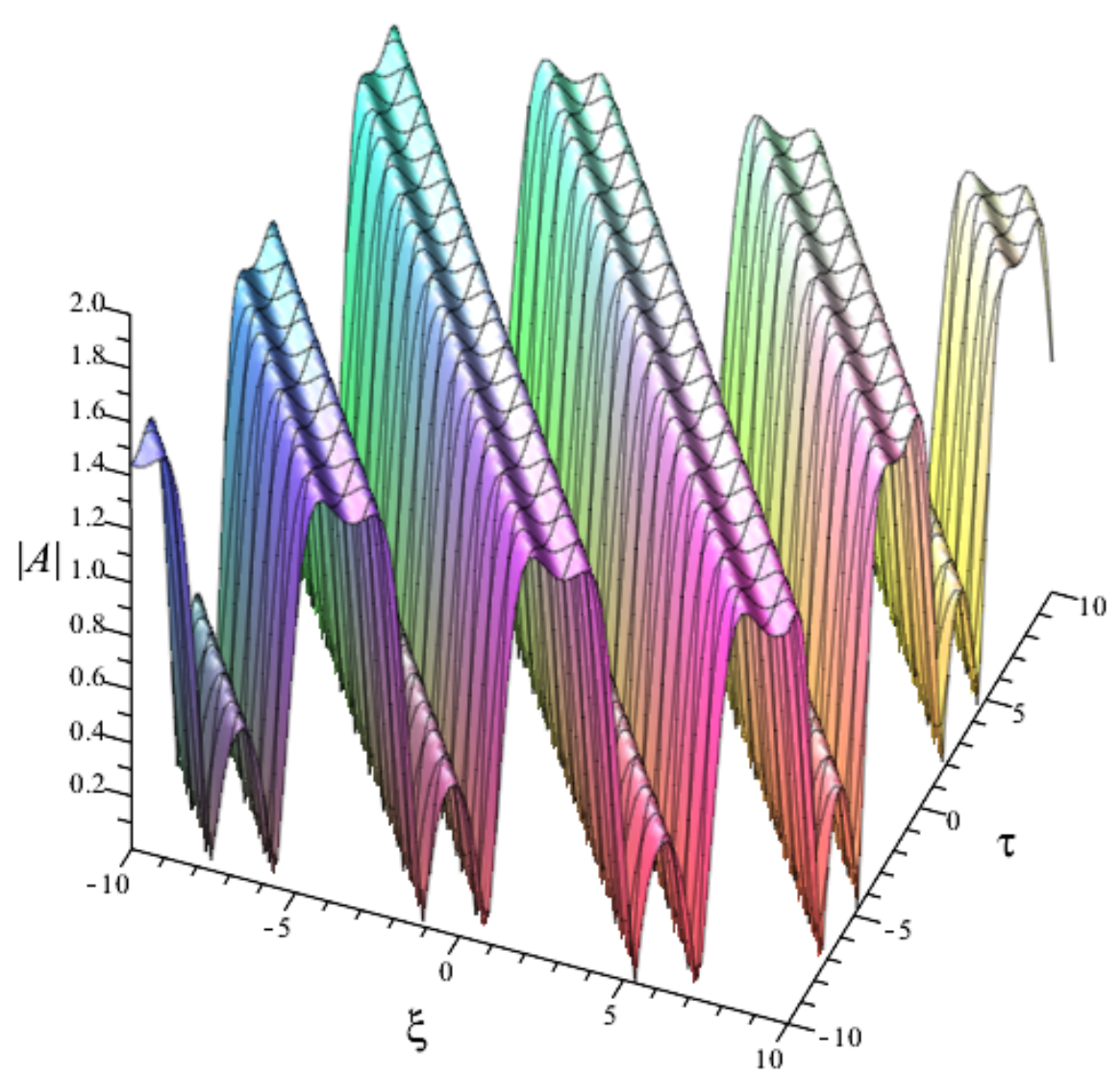}}
\subfigure[]{
\label{3dsol5b} %% label for first subfigure
\includegraphics[width=0.4\textwidth]{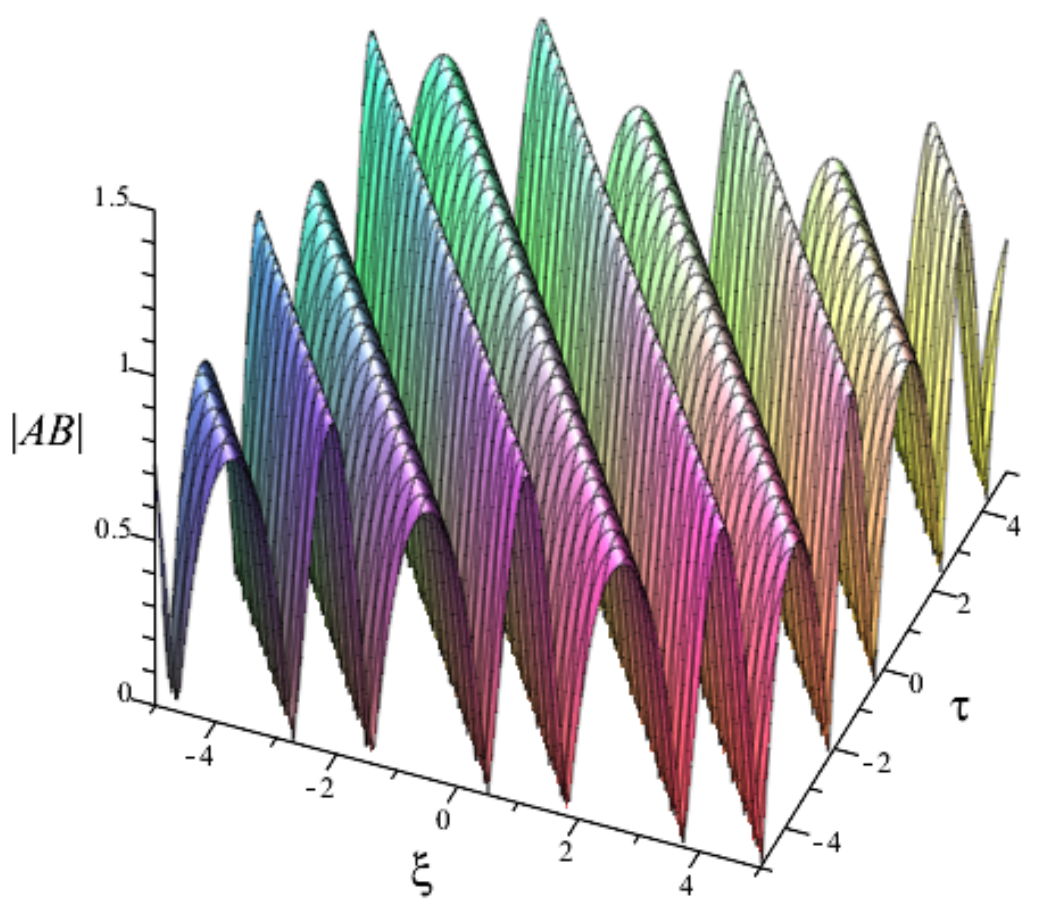}}
\caption{The three dimensional structure of the solution 5 with the parameters in Eq. \eqref{parasol4} for the quantity $A$ (left) and $AB$ (right).}
\label{3dsol5} %% label for entire figure
\end{figure}

\begin{figure}
\centering
\subfigure[]{
\label{1dsol5a} %% label for first subfigure
\includegraphics[width=0.4\textwidth]{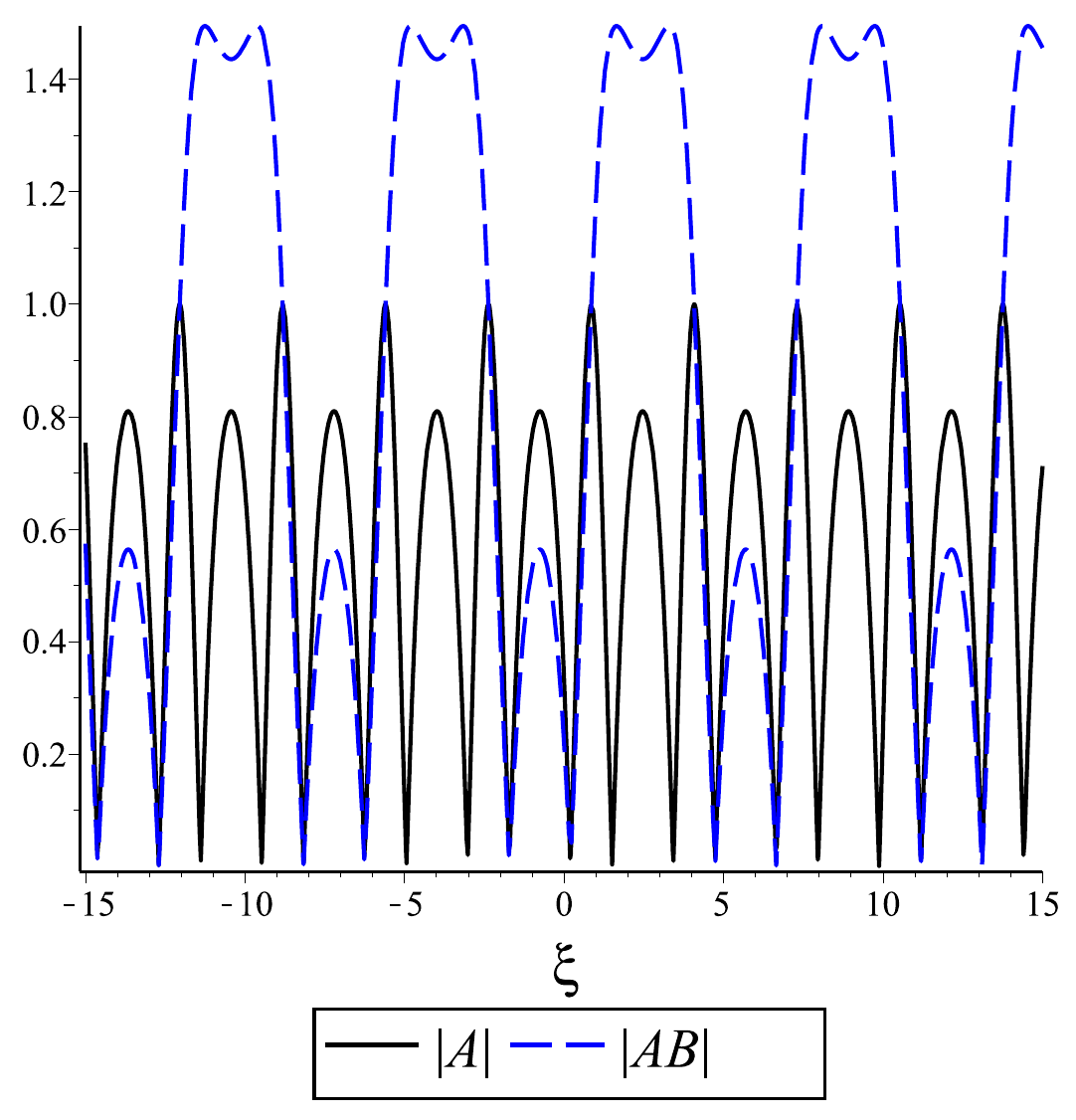}}
\subfigure[]{
\label{1dsol5b} %% label for first subfigure
\includegraphics[width=0.4\textwidth]{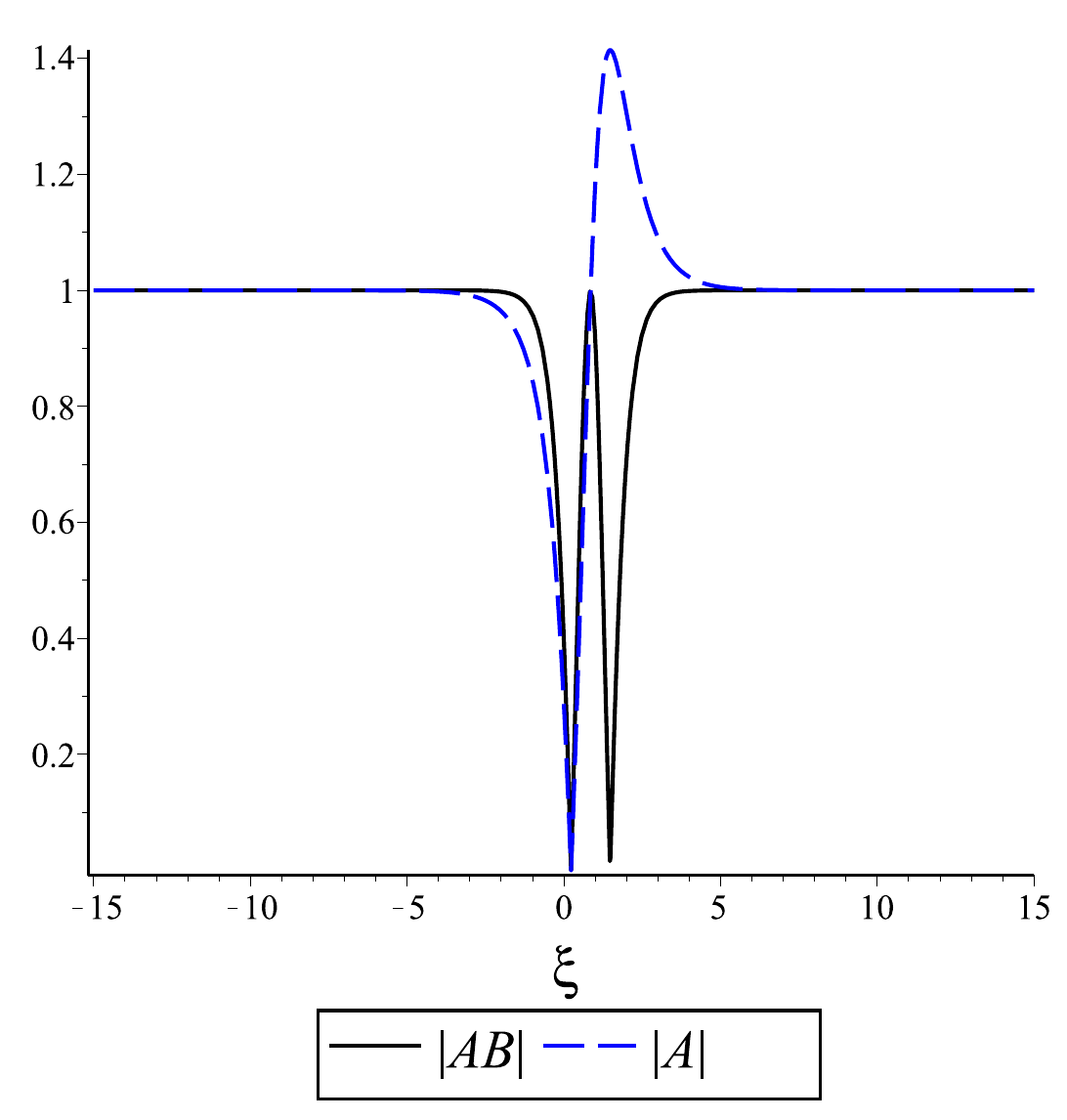}}
\caption{The one dimensional plot of the solution 5: (a) the wave patterns of $\abs{A}$ (blue-dash line) and $\abs{AB}$ (black-solid line) with the parameters in Eq. \eqref{parasol4} and $\tau=0$; (b) when the parameters being fixed as $e_1=e_2=e_4=e_5=e_6=k=\omega=\xi_0=\tau_0=1,\,e_3 = 0,\,e_7 = -1, m = 1$ the structure of $\abs{A}$ (blue-dash line) and $\abs{AB}$ (black-solid line).}
\label{1dsol5} %% label for entire figure
\end{figure}

Similarly, Fig. \ref{3dsol5a}, \ref{3dsol5b} show three dimensional structure of $\abs{A}$ and $\abs{AB}$ of the solution 5, respectively, with parameters being the same as \eqref{parasol4}, while Fig. \ref{1dsol5a} shows them both one-dimensionally at $\tau=0$. When taking $m=1$ and other parameters remain unchanged as in Eq. \eqref{parasol4}, the quantities $\abs{A}$ and $\abs{AB}$ of the solution 5 are depicted one dimensionally in Fig. \ref{1dsol5b}. It is interesting to find from Fig.  \eqref{3dsol5} that the quantity $\abs{A}$ represents a irregular wave, which have a small vibration on the top of the wave,  while the quantity $\abs{AB}$ represents a double-amplitude wave pattern, which can also be observed clearly from Fig. \ref{1dsol5a}. On the other hand, when taking $m=1$, Fig. \ref{1dsol5b} shows that the quantities $\abs{A}$ and $\abs{AB}$ have a dipole soliton structure, among which, $\abs{A}$ has one bright soliton and one dark soliton, while $\abs{AB}$ has two dark solitons.
\section{symmetry reduction solutions of the nonlocal nonlinear Schr\"{o}dinger equation}
In this section, we apply the standard Lie symmetry method to study the symmetry properties of the NNLS equation \eqref{nonnls} with \eqref{nonnlsB}. The general vector form of Lie point symmetry of the NNLS equation can be written as
\begin{equation}\label{vectorv1}
V=X\frac{\partial}{\partial \xi}+T\frac{\partial}{\partial
\tau}+\Gamma_1\frac{\partial}{\partial A}+\Gamma_2\frac{\partial}{\partial A^*}+\Lambda_1\frac{\partial}{\partial B}+\Lambda_2\frac{\partial}{\partial B^*}.
\end{equation}
In other words, the NNLS equation \eqref{nonnlsB} with its $\hat{P}_s^{\xi}\hat{T}_d^{\tau}\hat{C}$ counterpart
\begin{equation}\label{nonnls2}
iB_{\tau}+e_1B_{\xi\xi}+ie_2B_{\xi}+ie_3A_{\xi}+e_4\left|B\right|^2B+e_5A^*B^2+e_6(\left|B\right|^2A
+\left|A\right|^2B)+e_7A+e_7B=0,
\end{equation}
 are invariant under the transformation
\begin{equation}
\{\xi,\tau,A,A^*,B,B^*\} \rightarrow \{\xi+\epsilon X,\tau+\epsilon T,A+\epsilon \Gamma_1,A^*+\epsilon \Gamma_2,B+\epsilon \Lambda_1,B^*+\epsilon \Lambda_2\},
\end{equation}
with the infinitesimal parameter $\epsilon$.
Equivalently, the symmetry in the form \eqref{vectorv1} can be written as a function form as
\begin{subequations}\label{sigmasy}
\begin{equation}
\sigma_{A} = XA_{\xi}+TA_{\tau}-\Gamma_1,
\end{equation}
\begin{equation}
\sigma_{A^*} = XA^*_{\xi}+TA^*_{\tau}-\Gamma_2,
\end{equation}
\begin{equation}
\sigma_{B} = XB_{\xi}+TB_{\tau}-\Lambda_1,
\end{equation}
\begin{equation}
\sigma_{B^*} = XB^*_{\xi}+TB^*_{\tau}-\Lambda_2,
\end{equation}
\end{subequations}
where $\sigma_{A}$ and $\sigma_{B}$ satisfy the linearized equations of Eqs. \eqref{nonnls}, \eqref{nonnlsB} and \eqref{nonnls2}, i.e.,
\begin{subequations}\label{linear}
\begin{multline}
e_1\sigma_{A,\xi\xi}+i\sigma_{A,\tau}+ie_2\sigma_{A,\xi}+ie_3\sigma_{B,\xi}
+e_6(\sigma_{A^*}+\sigma_{B^*})A^2+e_6[(\sigma_{A^*}+\sigma_{B^*})B
\\+(A^*+B^*)(2\sigma_{A}+\sigma_{B})]A
+e_6\sigma_{A}(A^*+B^*)B+e_7(\sigma_{A}+\sigma_{B})=0,
\end{multline}
\begin{multline}
e_1\sigma_{B,\xi\xi}+i\sigma_{B,\tau}+ie_2\sigma_{B,\xi}+ie_3\sigma_{A,\xi}
+e_6(\sigma_{B^*}+\sigma_{A^*})B^2+e_6[(\sigma_{B^*}+\sigma_{A^*})B
\\+(A^*+B^*)\sigma_{B}]A
+e_6(\sigma_A+2\sigma_{B})(A^*+B^*)B+e_7(\sigma_{A}+\sigma_{B})=0,
\end{multline}
\begin{equation}
\sigma_{A^*}=\sigma_{B}(-\xi+\xi_0,-\tau+\tau_0),\,\sigma_{A}=\sigma_{B^*}(-\xi+\xi_0,-\tau+\tau_0),
\end{equation}
\end{subequations}

Substituting \eqref{sigmasy} with Eqs. \eqref{nonnls}, \eqref{nonnls2} into \eqref{linear} and vanishing all the coefficients of the independent partial derivatives of variables $A$, $A^*$, $B$ and $B^*$,  a system of over determined linear equations for $X, T, \Gamma_1, \Gamma_2, \Lambda_1, \Lambda_2$ are obtained. Calculated by computer, we get the desired solutions
\begin{eqnarray}\label{sol}
\nonumber&&X= \frac{c_1}{2}\xi,\,T=c_1\tau,\,\Lambda_1 =-\big[\frac{ic_1e_3e_7\xi+i e_7c_1(e_2-e_3)^2\tau }{e_2^2+e_3^2}-\frac{c_1}{2}\big]A\\\nonumber&& \Gamma_1 = \frac{ic_1e_2e_7\xi}{e_2^2+e_3^2}A+\big[\frac{ic_1e_3e_7\xi+ic_1e_7(e_2-e_3)^2\tau
}{e_2^2+e_3^2}-\frac{c_1}{2}\big]B,\\&&\Gamma_2=\Gamma_1^*,\,\Lambda_2=\Lambda_1^*,
\end{eqnarray}
under the condition of $e_4=e_5=e_6$ and $\xi_0=\tau_0=0$, where $c_1$ is an arbitrary constant.

Now, by substituting \eqref{sol} into  \eqref{sigmasy}, one obtains
\begin{subequations}\label{sigmaAB}
\begin{eqnarray}
\sigma_{A}&=&\frac{c_1}{2}\xi A_{\xi}+c_1\tau A_{\tau}-\frac{ic_1e_2e_7\xi}{e2^2+e3^2}A-\big[\frac{ic_1e_3e_7\xi+ic_1e_7(e_2-e_3)^2\tau
}{e2^2+e3^2}-\frac{c_1}{2}\big]B,\\
\sigma_{A^*}&=&\frac{c_1}{2}\xi A^*_{\xi}+c_1\tau A^*_{\tau}+\frac{ic_1e_2e_7\xi}{e2^2+e3^2}A^*+\big[\frac{ic_1e_3e_7\xi+ic_1e_7(e_2-e_3)^2\tau
}{e2^2+e3^2}+\frac{c_1}{2}\big]B^*,\\
\sigma_{B}&=&\frac{c_1}{2}B_{\xi}+c_1\tau  B_{\tau}+\big[\frac{ic_1e_3e_7\xi+i e_7c_1(e_2-e_3)^2\tau }{e_2^2+e_3^2}-\frac{c_1}{2}\big]A,\\
\sigma_{B^*}&=&\frac{c_1}{2}B^*_{\xi}+c_1\tau  B^*_{\tau}-\big[\frac{ic_1e_3e_7\xi+i e_7c_1(e_2-e_3)^2\tau }{e_2^2+e_3^2}+\frac{c_1}{2}\big]A^*.
\end{eqnarray}
\end{subequations}
The group invariant solutions of the NNLS equation can be obtained by solving \eqref{sigmaAB} under the condition $\sigma_A=\sigma_A^*=\sigma_B=\sigma_B^*=0$, alternatively, solving the corresponding characteristic equation
\begin{multline}\label{chac}
\frac{d\xi}{\frac{c_1}{2}\xi}=\frac{d\tau}{ c_1\tau}=\frac{dA}{ \frac{ic_1e_2e_7\xi}{e2^2+e3^2}A+\big[\frac{ic_1e_3e_7\xi+ic_1e_7(e_2-e_3)^2\tau
}{e2^2+e3^2}-\frac{c_1}{2}\big]B}=\frac{dB}{-\big[\frac{ic_1e_3e_7\xi+i e_7c_1(e_2-e_3)^2\tau }{e_2^2+e_3^2}-\frac{c_1}{2}\big]A}\\=\frac{dA^*}{-\frac{ic_1e_2e_7\xi}{e2^2+e3^2}A^*-\big[\frac{ic_1e_3e_7\xi+ic_1e_7(e_2-e_3)^2\tau
}{e2^2+e3^2}+\frac{c_1}{2}\big]B^*}=\frac{dB^*}{\big[\frac{ic_1e_3e_7\xi+i e_7c_1(e_2-e_3)^2\tau }{e_2^2+e_3^2}+\frac{c_1}{2}\big]A^*}.
\end{multline}

Under the condition $e_2=e_3$ and $2e_1e_7+e_3^2=0$, by solving out Eq. \eqref{chac}, the symmetry reduction solutions of the NNLS equation \eqref{nonnls} with \eqref{nonnlsB} are
\begin{eqnarray}\label{redA}
A&=&\frac{1}{\sqrt{\tau}}(\Phi_1\tau+\Psi_1 e^{\frac{2ie_7\sqrt{\tau}\zeta}{e_3}}),\\
\label{redB}B&=&-\frac{1}{\sqrt{\tau}}(\Phi_1\tau-\Psi_1 e^{\frac{2ie_7\sqrt{\tau}\zeta}{e_3}}),\\
\label{redA1}A^*&=&\sqrt{\frac{i}{\tau}}(\Phi_2\tau+\Psi_2 e^{\frac{-2ie_7\sqrt{\tau}\zeta}{e_3}}),\\
\label{redB1}B^*&=&-\sqrt{\frac{i}{\tau}}(\Phi_2\tau-\Psi_2 e^{\frac{-2ie_7\sqrt{\tau}\zeta}{e_3}}),
\end{eqnarray}
where $\Phi_1,\,\Psi_1,\,\Phi_2,\,\Psi_2$ are invariant functions of $\zeta=\frac{\xi}{\sqrt{\tau}}$.

Substituting Eqs. \eqref{redA}, \eqref{redB}, \eqref{redA1} and \eqref{redB1} into the NNLS equation \eqref{nonnls}, \eqref{nonnls2} with \eqref{nonnlsB} yields the symmetry reduction equations
\begin{equation}\label{redefs1}
4e_6e_7(1+i)\sqrt{2}\Psi_2\Psi_1^2+ie_7\Psi_1+i e_7\Psi_{1,\zeta}\zeta+e_3^2\Psi_{1,\zeta\zeta}=0,
\end{equation}
\begin{equation}\label{redefs2}
4e_6e_7(1+i)\sqrt{2}\Psi_2\Phi_1\Psi_1-ie_7\Phi_1+i e_7\Phi_{1,\zeta}\zeta+e_3^2\Phi_{1,\zeta\zeta}=0,
\end{equation}
\begin{equation}\label{redAB}
\Psi_2(\zeta)=\frac{1+i}{2}\sqrt{2}\Psi_1(i\zeta),\,\Phi_2(\zeta)=\frac{1+i}{2}\sqrt{2}\Phi_1(i\zeta),
\end{equation}
with $\Psi_1(-\zeta)=-\Psi_1(\zeta)$ and $\Phi_1(-\zeta)=-\Phi_1(\zeta)$.

\section{Conclusion and discussion}
In summary, a general form of NNLS equation with shifted space parity, delayed time reversal and charge-conjugate is derived from a nonlocal version of two-layer liquid model by using multi-scale expansion method. Various types of periodic wave solutions of the NNLS equation are obtained by using elliptic expansion method, which become kink (anti-kink) or bright (dark) type soliton solutions when the modulus approaching to unity. Also, some figures are plotted and analyzed to reveal some interesting features of these solutions. Applying Lie's standard symmetry method directly to nonlocal systems is a challenging task, which are rarely investigated in other literatures. Fortunately, we successfully obtained not only the symmetry group but also symmetry reduction solution under some constraints of the coefficients of the NNLS equation, through which plenty of new solutions could be found. In consideration of the important role of the nonlinear Schr\"{o}dinger equation playing in various physical fields, it is deserved to be studied furthermore in the future.

\begin{acknowledgments}
 This work was supported by the National Natural Science Foundation of China under Grant Nos. 11405110, 11275129 and the Natural Science Foundation of Zhejiang Province of China under Grant No. LY18A050001.
\end{acknowledgments}
\section*{Compliance with ethical standards}
\section*{Conflict of interest statement}
The authors declare that they have no conflicts of interest to this work. There is no professional or other personal interest of any nature or kind in any product that could be construed as influencing the position presented in the manuscript entitled.

\end{document}